\documentclass[12pt]{iopart}

\begin{document}

\title[Zhu $et$  $al$., Nernst effect of LaO$_{1-x}$F$_{x}$FeAs]{Nernst effect of the new iron-based superconductor LaO$_{1-x}$F$_{x}$FeAs}

\author{Z W Zhu$^{1}$, Z A Xu$^{1}$\footnote[3]{Corresponding
author. Tel: (86)571-87953255, E-mail address: zhuan@zju.edu.cn},
X Lin$^{1}$, G H Cao$^{1}$, C M Feng$^{2}$, G F Chen$^{3}$, Z
Li$^{3}$, J L Luo$^{3}$, N L Wang$^{3}$}

\address{$^{1}$ Department of Physics, Zhejiang Unviersity, Hangzhou 310027, China}
\address{$^{2}$ Test and Analysis Center, Zhejiang University, Hangzhou 310027, China}
\address{$^{3}$ Beijing National Laboratory for Condensed Matter Physics, Institute of Physics, Chinese Academy of Sciences, Beijing 100080, China}
\begin{abstract}
We report the first Nernst effect measurement on the new
iron-based superconductor LaO$_{1-x}$F$_{x}$FeAs $(x=0.1)$. In the
normal state, the Nernst signal is negative and very small. Below
$T_{c}$ a large positive peak caused by vortex motion is observed.
The flux flowing regime is quite large compared to conventional
type-II superconductors. The sharp decrease in Nernst signal at
the depairing magnetic field $H_{c2}$ is not obvious even for
temperatures close to $T_c$. Furthermore, a clear deviation of the
Nernst signal from normal state background and an anomalous
suppression of off-diagonal thermoelectric current in the normal
state between $T_{c}$ and $50$ K are observed. We argue that this
anomaly in the normal state Nernst effect could correlate with the
SDW fluctuations although the contribution of superconducting
fluctuations can not be excluded.

\end{abstract}

\maketitle

\section{Introduction}
Since the discovery of superconductivity in the quaternary
oxy-pnictides LaOFeP \cite{Jap1} and LaONiP \cite{Jap2}, a new
family of high-$T_{c}$ superconductors featured with iron-based
layered structure are emerging \cite{Jap3,4NLWang,Wen,
Wen5,ChenXianhui6,RenZA7,RenZA7a,Wang8}. Non-conventional
superconducting mechanism has been proposed based on the tunneling
spectroscopy\cite{Wen9}, thermodynamics measurements\cite{Wen10},
and the first-principle calculation
\cite{11FangZhong,12Luzhongyi}. The recent band structure
calculation indicates the Fermi surface (FS) of
LaO$_{1-x}$F$_{x}$FeAs has five sheets
\cite{11FangZhong,12Luzhongyi,13arXiv}: two cylinders associated
with electron type around the zone edge $M$-$A$ line of the
Brillouin zone, two hole-like cylinders along the $\Gamma$-$Z$
line, and a hole-like closed pocket centered at $Z$ point. The
calculations also indicate that LaOFeAs is basically characterized
as low density electron carriers doped on top of a band insulator
with filled d$^6$ valence orbitals \cite{11FangZhong}. F doping
can reduce the hole-like sheet and enhance the electron-like FS.
Hall effect measurements\cite{4NLWang,Wen} have confirmed the
electron-type conductivity with a charge carrier density as low as
$1.8\times10^{-21}$cm$^{-3}$ for F doped systems, even lower than
that of high-$T_{c}$ cuprates.

The low dimensionality and low charge carrier density in this
system remind us of the analogy to high-$T_{c}$ cuprates. Hence
the superfluid density $n_s$ in this system should be also very
small compared to typical metal or alloy superconductors. It has
been proposed that the small superfluid density results in a small
phase stiffness energy and the Meissner state is readily destroyed
by the strong phase fluctuations \cite{14Emery}. In the case of
high-$T_{c}$ cuprates, the correlation between the $T_{c}$ and
superfluid density demonstrated by the so-called "Uemura plot"
\cite{15Uemura} strongly implies such a phase disorder scenario in
which the superconducting transition at $T_{c}$ is driven by the
loss of long-range phase coherence rather than "gap closing".
Furthermore, persistence of enhanced Nernst signals due to the
vortex excitations at temperatures rather high above $T_{c}$
provides strong evidence for the spontaneous appearance of vortex
excitations above $T_{c}$\cite{XuNature,OngReview}.

The Nernst effect is a magneto-thermoelectric effect, defined as
the appearance of a transverse electric field $E_{y}$ in response
to a temperature gradient $\nabla T || x$, in the presence of a
perpendicular magnetic field $H||z$ and under open circuit
conditions. The Nernst effect is usually small in usual metals
 because of the so-called "Sondheimer
cancellation"\cite{16Cancellation}. It is known that the magnetic
ordering and spin fluctuations can cause enhanced Nernst signals
\cite{17WLLi,Kon}. For a type-II superconductor (in the
vortex-liquid state), vortices can be driven down by the
temperature gradient with velocity $v$ along the $x$-axis
direction. Moving vortices can produce a transverse electric field
according to the Josephson equation: $E_{y}=B\times v$ , which is
also called vortex Nernst effect. Thus the vortex Nernst effect is
a highly sensitive probe to detecting vortex motion.

Here we report the first Nernst effect measurement on the
newly-discovered iron-based superconductor LaO$_{1-x}$F$_{x}$FeAs$
(x=0.1)$. A broad regime below $T_{c}$ featured by large positive
vortex Nernst signal is observed, which indicates that the flux
flowing regime is quite large compared to conventional type-II
superconductors. However, a clear anomalous depression of
off-diagonal thermoelectric current below $50$ K in the normal
state is observed and its origin is discussed.

\section{Experimental}
The polycrystalline samples were prepared by the solid state
reaction using LaAs, Fe$_{2}$O$_{3}$, Fe and LaF$_{3}$ as starting
materials. The sample preparation details can be found in Ref.[4].
The powder X-ray diffraction patterns indicate that the resultant
is single phase and all the diffraction peaks can be well indexed
based on the tetragonal ZrCuSiAs-type structure with the space
group P4/nmm. The structure characterization was reported in Ref.
[4]. The nominal composition is LaO$_{0.9}$F$_{0.1}$FeAs according
to the starting materials. The resistivity measurement shows a
sharp superconducting transition and $T_{c}$ (the onset
temperature in the resistive transition) is $26$ K. The residual
resistance ratio (RRR) which is defined as $R$(300 K)/$R$(28 K) is
11.6, indicating a good sample quality.

The Hall effect was measured by scanning magnetic field at fixed
temperatures. The thermoelectric properties were measured by a
steady-state technique. The temperature gradient used for the
thermoelectric measurements, determined by a pair of differential
Type E thermocouples, was around $0.5$ K/mm. All the measurements
were performed in a Quantum Design PPMS-$9$ system. The Nernst
signal $e_{y}$ is defined as $e_{y}\equiv\frac{E_{y}}{|\nabla T|}$
. Nernst signal was measured at positive and negative field
polarities, and the difference of the two polarities was taken to
remove any thermopower contribution.

\section{Results and discussion}

Traces of Nernst signal as a function of magnetic field up to
$\mu_0$$H$ of 8 T is displayed in Figure 1 at various
temperatures. Above $T_{c}$ of about $26$ K, the Nernst signal
($e_{y}$) is negative, small, and linear as a function of magnetic
field. Below $T_{c}$, the magnitude of the Nernst signal increases
sharply, and becomes nonlinear with magnetic field, which exhibits
a typical feature of vortex Nernst signal. At $10$ K, vortices
form a solid lattice and are pinned so that the vortex Nernst
signal is nearly zero. At $T =13$ and $15$ K, the vortex lattice
melts at a certain magnetic field $H_{m}$ and vortex Nernst
signals appears for $H > H_{m}$. For $T > 17$ K, such a melting
transition of vortex lattice is no longer observed just because
solid vortex lattice does not form due to thermal fluctuations.
Probably because of the polycrystalline sample, the vortex motion
is not uniform and enhanced noises are observed which might be
caused by the vortex jumping. It is interesting that the traces of
the vortex Nernst signals below $T_{c}$ are similar to these of
high-$T_{c}$ cuprates, i.e., the vortex liquid state regime is
quite large and the solid vortex lattice is hard to form. Another
surprising feature is that the depairing magnetic field $H_{c2}$
cannot be defined in the $e_{y}$ vs $H$ curves even for
temperatures near $T_{c}$. The magnetoresistance measurement
performed on the same batch of LaO$_{1-x}$F$_x$FeAs ($x$=0.1)
samples  was reported in Ref.[4] and $H_{c2}(T)$ was determined
from the onset transition of resistance for temperatures close to
$T_c$. $\mu_0H_{c2}(0)$ was estimated to be about 55 T
 and $H_{c2}(T)$ drops down to zero quickly when $T$
approaches $T_{c0}$. However, the Nernst effect study indicates
that $H_{c2}$ might not drop to zero at $T_{c}$ and
superconducting transition could be driven by the loss of
long-range phase coherence of Cooper pairs, just as the case in
the high-$T_{c}$ superconducting cuprates \cite{18WangOng}. Such a
behavior is contrast to conventional layered superconductors such
as NbSe$_{2}$\cite{19Ong,20Behnia}.

The temperature dependence of Nernst coefficient, $\nu(T)$,
together with its resistivity, $\rho(T)$, is shown in Figure 2.
The Nernst coefficient is defined as the initial slope of
$e_{y}$-$H$ curves for $T<T_{c}$. The normal state Nernst
coefficient is very small and shows weak temperature dependence.
However $\nu$ starts to increase below 50 K, change sign just
above $T_{c}$, and then shows a step-like sharp increase at
$T_{c}$. Below $T_{c}$, it reaches a peak around 15 K. The peak
value is around 0.10$\mu$V/KT, smaller than vortex Nernst signal
observed in high- $T_{c}$ superconductors, probably because of the
polycrystalline sample in this study. The sharp increase in $
\nu(T)$ occurrs at $T_{c}$, in contrast to the smooth change of
$\nu(T)$ at $T_{c}$ in high- $T_{c}$ superconductors. However, for
$T$ as high as 50 K above $T_{c}$, $\nu(T)$ already starts to
deviate from the negative background and increase gently. In high-
$T_{c}$ cuprates, a similar enhancement of Nernst signal has
already been observed and it has been interpreted as the evidence
for vortex excitations above $T_{c}$. A phase disorder scenario
based on the Nernst effect measurements is proposed for
high-$T_{c}$ cuprates \cite{XuNature,OngReview}. Because this
iron-based new superconductors share the similarity in the low
dimensionality and low superfluid density with high- $T_{c}$
cuprates, it is plausible to ascribe the deviation of Nernst
signal from normal state background to the vortex excitations
existing in a region up to 20 K above $T_{c}$. However, as
discussing whereafter, we argue that this anomaly change in Nernst
signal between $T_{c}$ and 50 K might correlate with the
spin-density wave (SDW) fluctuations rather than superconducting
phase fluctuations.

In order to explore the nature underlying this anomaly in the
Nernst effect, it is useful to isolate the off-diagonal
thermoelectric (Peltier conductivity) term from the Nernst signal
by measuring the Hall effect and Seebeck effect separately. The
temperature dependence of Hall coefficient ($R_{H}$) and
thermopower ($S$) was measured for the same sample under magnetic
field ($\mu_0H$) of 8 T, and they are shown in Figure 3(a)and (b).
Their values in the normal state are consistent with previous
reports \cite{4NLWang,Wen,23Oak} and the negative signs mean that
the dominant charge carrier is electron-like. At $T_{c}$, both
$R_{H}$ and $S$ drop to zero sharply. It is very interesting that
the profile of $S$ vs $T$ curves is similar to that of low charge
density metals like underdoped high-$T_{c}$ cuprates and
Na$_{x}$CoO$_{2}$ except the negative sign.

The normal state Nernst signal is comprised of two terms, viz.
\begin{equation}
e_{y}=\rho\alpha_{xy}-S\tan\theta,
\end{equation}
where $\alpha_{xy}$ is the off-diagonal Peltier coefficient, $S =
\frac{\alpha}{\sigma}$ the thermopower ( $\sigma$ the diagonal
conductivity and $\alpha$ the diagonal Peltier coefficient),
$\rho$ the resisitivity, and tan$\theta$  the Hall angle. For
usual simple metals, the two terms in Eq. (1) cancel each other,
which is so-called "Sondheimer cancellation"
\cite{16Cancellation}. The second term $S$tan$\theta$ is
calculated based on the $S$ and Hall effect data, which is shown
in Figure 4(a). By comparing $e_{y}$ with $S$tan$\theta$  in
Figure 4(a), it becomes clear that the Nernst signal $e_{y}$ is
about one fifth of $S$tan$\theta$ , which means the "Sondheimer
cancellation" is not complete\cite{24Note}. However the multiband
character does not enhance significantly the Nernst signals in
normal state. In other words, only one type (electron type) charge
carrier is dominant in the Nernst effect of the F doped
LaO$_{1-x}$F$_{x}$FeAs. The first term $\rho\alpha_{xy}$ in Eq.(1)
is calculated by subtracting $e_{y}$ from the product
$-S$tan$\theta$ , which is shown in Figure 4(b). It can be found
that the off-diagonal thermoelectric current $\alpha_{xy}$
increases with decreasing temperature at high temperatures, but it
saturates around 50 K and starts to decrease quickly as the
temperature is close to $T_{c}$. Around and below $T_{c}$, there
are three terms which comprise $e_{y}$, viz.,
\begin{equation}
e_{y}=\rho\alpha^{s}_{xy}+\rho\alpha^{n}_{xy}-S\tan\theta,
\end{equation}
where $\rho\alpha_{xy}^{s}$ denotes the contribution of vortex to
the Nernst signals, and $\rho\alpha_{xy}^{n}$ denotes the
contribution from normal state excitations. Below $T_{c}$,
$\rho\alpha_{xy}^{n}$ should drop to zero quickly, and
$\rho\alpha_{xy}^{s}$ shows a large positive peak. However,
$\rho\alpha_{xy}^{n}$ term starts to decreases at temperatures far
above $T_{c}$, which means a change in the electron state below 50
K. The off-diagonal Peltier coefficient ($\alpha_{xy}$) shows a
similar profile as $\rho\alpha_{xy}^{n}$. It has been proposed
that the undoped parent compound LaOFeAs undergoes a SDW
transition around 134 K and the F doping will suppress the SDW
transition, and finally give a way to superconductivity
\cite{25Dong,26Cruz, 27McGuire}. In F doped superconductors, there
are evidences to support the existence of magnetic fluctuations.
The variation of the temperature dependence of the resistivity
with F content shows that SDW transition is gradually suppressed
by F doping, and competition of SDW order and superconductivity
order was proposed \cite{25Dong}. Recently Hosono and his
collaborators reported that a pseudogap of 0.1 eV in
LaO$_{1-x}$F$_{x}$FeAs was detected by photoemission and they
attributed the origin of the pseudogap to strong spin fluctuations
\cite{31Hosono}. They also found that the spin fluctuation shows a
maximum at 5\% F doping by DC magnetization measurements
\cite{32Hosono}.  The slow increase in $e_{y}$ above 50 K might be
caused by the residual magnetic fluctuations, and therefore we
suggest that the sharp decrease in $\rho\alpha_{xy}^{n}$ below 50
K could result from the suppression of magnetic fluctuations. In
other words, there is a kind of "precursor state" between $T_{c}$
and 50 K in which the magnetic fluctuations are strongly
suppressed. Such a suppression of magnetic fluctuations should be
detected by other probes like nuclear magnetic resonance (NMR)
measurement. Recent NMR measurements by Nakai \etal
\cite{33Hosono} indicated that the 1/$TT_1$ of $^{75}$As in the
LaO$_{0.96}$F$_{0.4}$FeAs sample shows a Cuire-Weiss behavior down
to 30 K, suggesting the development of SDW spin fluctuations with
decreasing temperatures. Then it is suppressed below 30 K. They
concluded that the superconductivity emerges when a magnetic
ordering is suppressed. We suggest that the sharp suppression of
the off-diagonal Peltier current $\alpha_{xy}$ and the suppression
of 1/$TT_1$ just above $T_c$ should have the same origin, i.e.,
both should result from the suppression of SDW fluctuations. It is
an interesting open question whether this "precursor state" is
indeed crucial to the emergence of "high-$T_{c}$"
superconductivity in this system.

\section{Conclusion}

We report the first Nernst effect measurement on the new
iron-based superconductor LaO$_{1-x}$F$_{x}$FeAs ($x$=0.1). In the
normal state, the Nernst signal is negative and very small, which
means that the "Sondheimer cancellation" is incomplete and the
normal state transport is mainly dominate by the electron-like
band. Below $T_{c}$ a large positive peak related to vortex motion
is observed. The flux flowing regime is very large, and there is
no obvious change in Nernst signals at $H_{c2}$ which is defined
from the magnetoresistance, contrast to conventional type-II
superconductors with layered structure. Furthermore, a clear
deviation of the Nernst signal from normal state background and an
anomalous suppression of off-diagonal thermoelectric current
between $T_{c}$ and 50 K are observed. We argue that this anomaly
in the normal state Nernst effect could correlate with the SDW
fluctuations.

\section*{Acknowledgments}
We thank F. C. Zhang, Z. Y. Weng, H. H. Wen, X. H. Chen, and T.
Xiang for helpful discussions. This project is supported by the
National Science Foundation of China and the National Basic
Research Program of China (Grant No. 2006CB601003 and 2007CB925001
). ZAX and GHC also acknowledge the support by PCSIRT (Grant No.
IRT0754).

\pagebreak[4]

\begin{figure}
\caption{\label{Fig1}(Color Online) Magnetic field dependence of
the Nernst signal ($e_{y}$) at different temperatures.}
\end{figure}

\begin{figure}
\caption{\label{Fig2} (Color Online) Temperature dependence of the
Nernst coefficient. The resistivity measured under zero magnetic
field is shown together. The arrow indicates $T_{c}$ of 26 K. The
dashed line is a guide to the eye.}
\end{figure}

\begin{figure}
\caption{\label{Fig3} Panel (a): Temperature dependence of the
Hall coefficient ($R_{H}$) measured under magnetic field
($\mu_0H$) of 8 T. Panel (b): Temperature dependence of
thermopower ($S$). The inset shows the enlarged plot of $S$ versus
$T$ around $T_{c}$.}
\end{figure}

\begin{figure}
\caption{\label{Fig4} (Color Online) Panel (a) Temperature
dependence of the Nernst signal ($e_{y}$), the product term
-$S$tan$\theta$, and off-diagonal Peltier term $\rho\alpha_{xy}$.
Panel (b): Temperature dependence of the off-diagonal Peltier
coefficient $\alpha_{xy}$. The arrow indicates $T_{c}$ of 26 K.}
\end{figure}


\begin{thebibliography}{10}
\bibitem{24Note}In the case of hole-doped high-$T_c$ cuprates, $e_y$ is only about 1/10 of $S$tan$\theta$,  i.e., the Sondheimer cancellation is nearly complete.
\end{thebibliography}
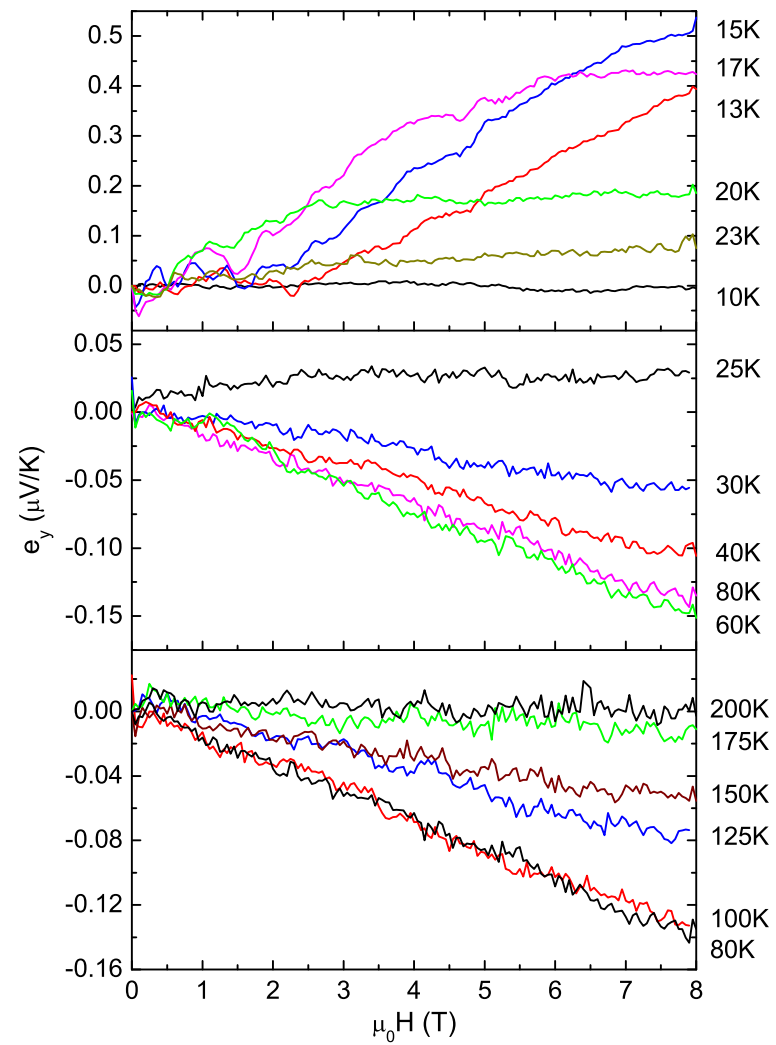
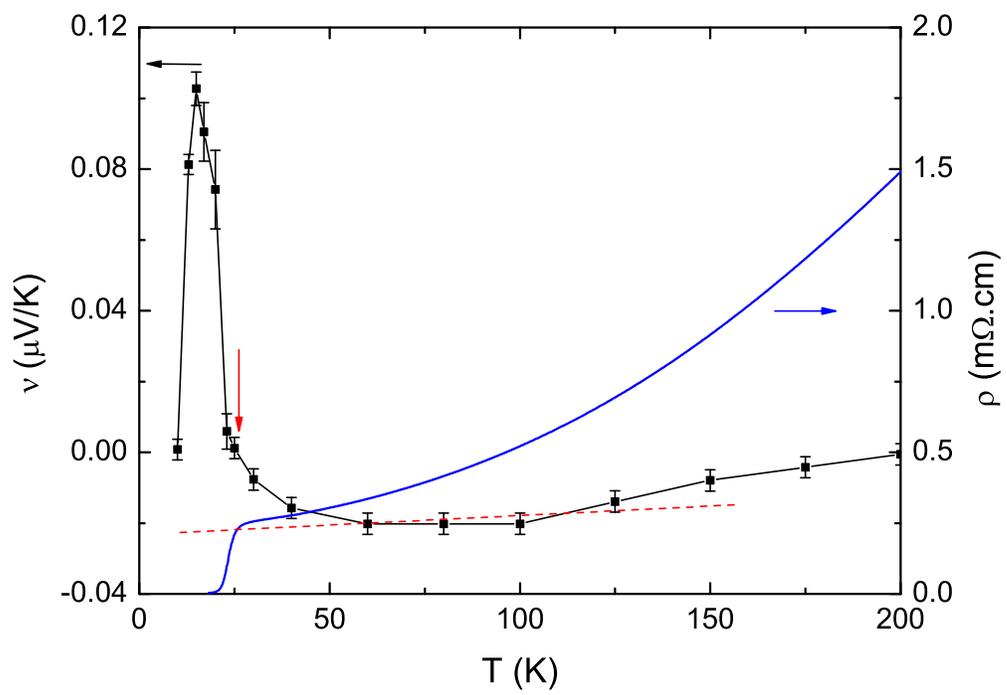
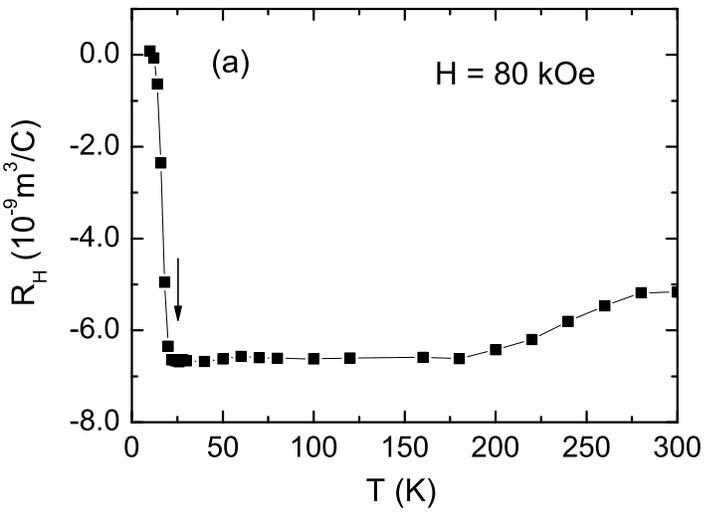
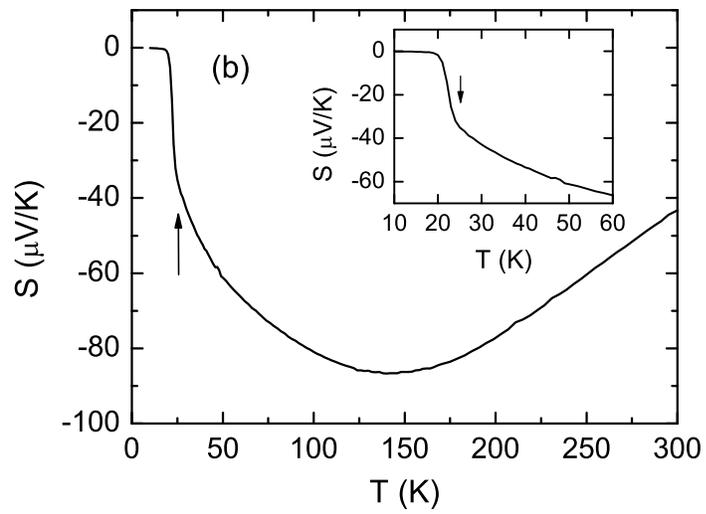
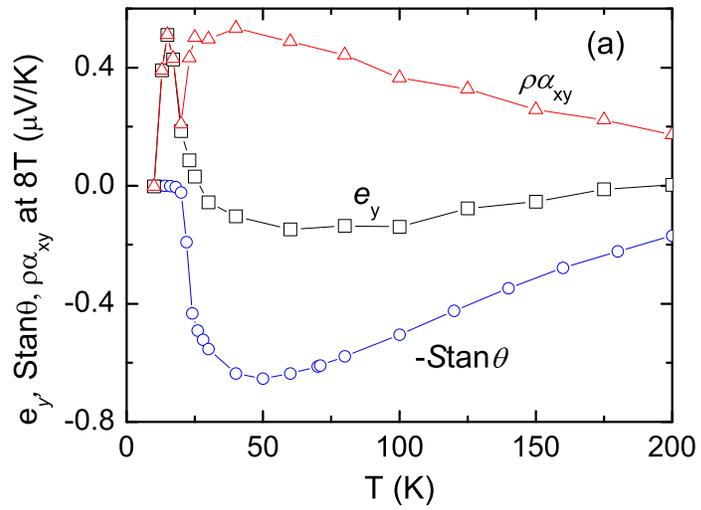
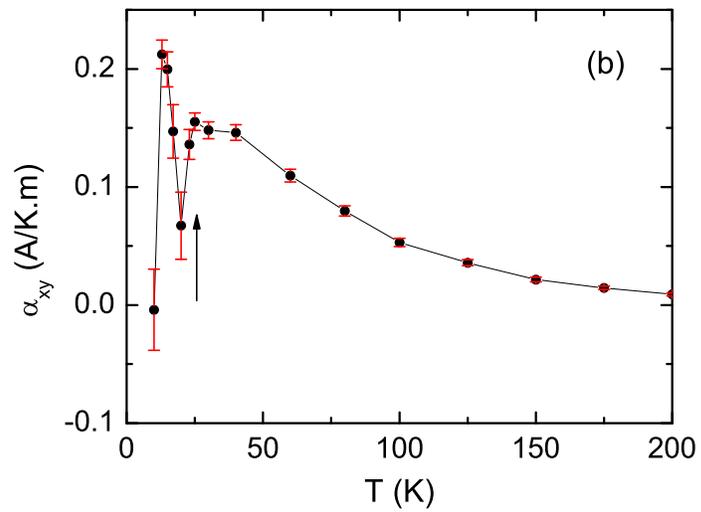
\end{document}